\documentclass{jnmp}

\usepackage{amsmath}

\JNMPnumberwithin{equation}{section}

\newtheorem{theorem}{Theorem}

\newtheorem{proposition}{Proposition}
\theoremstyle{definition}

\newtheorem{remark}{Remark}

\newcommand{\bbR}{\mathbb{R}}
\newcommand{\bbZ}{\mathbb{Z}}
\newcommand{\bbN}{\mathbb{N}}

\newcommand{\bbC}{\mathbb{C}}

\newcommand{\ad}{\mathrm{ad}}

\newcommand{\cD}{{\mathcal{D}}}
\newcommand{\cP}{{\mathcal{P}}}
\newcommand{\cK}{{\mathcal{K}}}

\newcommand{\cO}{{\mathcal{O}}}

\newcommand{\cS}{{\mathcal{S}}}

\newcommand{\Vect}{\mathrm{Vect}}
\newcommand{\vol}{\mathrm{vol}}

\newcommand{\half}{\frac{1}{2}}

\def\d{\delta}

\def\b{\beta}

\begin{document}

%
\renewcommand{\evenhead}{N Ben Fraj and S Omri}
\renewcommand{\oddhead}{Deforming the Lie Superalgebra of Contact
Vector Fields on $S^{1|1}$ }

%
\thispagestyle{empty}

\FirstPageHead{*}{*}{20**}{\pageref{firstpage}--\pageref{lastpage}}{Article}

\copyrightnote{2005}{N Ben Fraj and S Omri}

\Name{Deforming the Lie Superalgebra of Contact
Vector Fields on $S^{1|1}$ inside the Lie Superalgebra of
Superpseudodifferential operators on $S^{1|1}$ }

\label{firstpage}

\Author{N BEN FRAJ~$^\dag$ and S Omri~$^\ddag$}

\Address{$^\dag$ Institut Sup\'{e}rieur de Sciences Appliqu\'{e}es
et Technologie, Sousse, Tunisie \\
~~E-mail:~benfraj\_nizar@yahoo.fr\\[10pt]
$^\ddag$  D\'epartement de Math\'ematiques, Facult\'e des Sciences de Sfax,
Route de Soukra,\\~~3018 Sfax BP 802, Tunisie\\
~~E-mail:~omri\_salem@yahoo.fr}

\Date{Received Month *, 200*; Revised Month *, 200*;
Accepted Month *, 200*}

\begin{abstract}
\noindent
We classify nontrivial deformations of the standard embedding of
the Lie superalgebra K(1) of contact vector fields on the (1,1)-dimensional
supercircle into the Lie superalgebra
of superpseudodifferential operators on the supercircle. This approach
leads to the deformations of the central charge induced on
K(1) by the canonical central extension of $S\Psi DO$.
\end{abstract}


\section{Introduction}
The study of multi-parameter deformations of the standard
embedding of the Lie algebra $\Vect(S^1)$ of vector fields on the
circle $S^1$ inside the Lie algebra $\Psi \cD O$ of
pseudodifferential operators on $S^1$ was carried out in
\cite{OR1, OR2}. In this paper we address the
computation of the integrability conditions of infinitesimal
deformations of the standard embedding of the Lie superalgebra
$\cK(1)$ of contact vector fields on the supercircle $S^{1|1}$
inside the Lie superalgebra $\cS \Psi \cD \cO$ of
superpseudodifferential operators on $S^{1|1}$. The infinitesimal
deformations of this embedding are classified by $H^1(\cK(1),\cS \Psi \cD \cO)$. This space is
four dimensional and it was  calculated in \cite{AB}. The
obstructions for integrability of infinitesimal deformations
lie in
$H^2(\cK(1),\cS \Psi \cD \cO)$. Our goal is to study these
obstructions.

It turns out that there exist four even
one-parameter families of nontrivial deformations. We will compute
explicit formulas describing these families. A contraction
procedure of those deformations leads to four one-parameter
deformations of the standard embedding of $\cK(1)$ into the
Poisson Lie superalgebra $\cS\cP$ of superpseudodifferential
symbols on $S^{1|1}$. Each parameter describes an interesting
algebraic curve in the space of parameters.

The well-known nontrivial central extension  of $\cS \Psi \cD
\cO$ induces a central extension of the subsuperalgebra
$\cK(1)$(see \cite{AOR1}). The restriction of the 2-cocycle
generating this extension is the 2-cocycle defining the central
extension of $\cK(1)$ known as the Neveu-Schwartz Lie superalgebra
(see \cite{AOR1}, \cite{MR}). As an application of our results, we
obtain a \lq \lq deformed" expression for the central charge
induced by the deformations of the standard embedding we have
constructed.


\section{ The main definitions}

\subsection{Superpseudodifferential operators on $S^{1|1}$}
We first recall the definition of $\cS \Psi \cD \cO$ (cf. \cite
{CMZ, MR}). The supercircle $S^{1\,|1}$ is the superextension of
the circle $S^1$ with local coordinates $(x,\theta)$, where $x\in
S^1$ and $\theta$ is odd. A $C^{\infty}$-function on $S^{1|1}$ has
the form $F=f(x)+2g(x)\theta$, with $f,g\in C^{\infty}(S^1)$. The
vector field $\displaystyle \eta=\frac{\partial}{\partial
\theta}+\theta \frac{\partial}{\partial x}$ on $S^{1|1}$ sends $F$
to $\eta(F)=2g(x)+f'(x)\theta$, so that $\displaystyle\eta^2=\half
[\eta,\eta] =\frac{\partial}{\partial x}$. The usual Leibniz rule:
$\displaystyle \frac{\partial}{\partial x}\circ
f=f'(x)+f(x)\frac{\partial}{\partial x}$  on $ C^{\infty}(S^1)$,
is replaced on $C^{\infty}(S^{1|1})$ by ($p$ is for parity):
\begin{equation}
\label{leibniz}
 \eta\circ F=\eta(F)+\sigma(F)\eta, \;\text{ where $\sigma(F)=(-1)^{p(F)}F$}.
\end{equation}

Formula (\ref{leibniz}) generalizes by induction on $m$ to the
graded Leibniz formula:
\begin{equation}
\label{gleibniz} \eta^m\circ
F=\displaystyle\sum_{k=0}^{\infty}(^m_k)_s\eta^k(\sigma^{m-k}(F))\eta^{m-k}
\end{equation}
for all integer $m$, where the super binomial coefficients
$(^m_k)_s$ are defined by:
\begin{equation*}
 (^m_k)_s=
\begin{cases}
\binom{[\frac{m}{2}]}{[\frac{k}{2}]}&\text{if either $k$ is even or $m$ is odd }\\
~~0 &\text{otherwise},
\end{cases}
\end{equation*}
$[x]$ is the integer part of a real
number $x$, and $(^x_l)=\frac {x(x-1)\cdots (x-l+1)}{\ell!} $ for $l\in \bbZ_{\geq 0}$.
Set:
\begin{equation*} \cS \Psi\cD \cO=\Biggl\{\sum_{k\in
\bbZ_{\geq 0}} F_k \;\eta^{\omega-k}\;\mid\;  w\in \bbZ, \; F_k\in
C^{\infty}(S^{1|1})\Biggr\},
\end{equation*}
where the composition of
superpseudodifferential operators is given by (\ref{gleibniz}):
\begin{equation*}
\label{product}
 F\eta^m \circ
G\eta^n=\displaystyle\sum_{k=0}^{\infty}(^m_k)_s F
\eta^k(\sigma^{m-k}(G))\eta^{m+n-k}~~\makebox{for any}~~ m,n\in
\bbZ \makebox{ and } F,\, G\in C^{\infty}(S^{1|1}).
\end{equation*}
Denote by $\cS\Psi\cD \cO_{SL}$ the Lie superalgebra with the same
 superspace as $\cS\Psi\cD \cO$ and the supercommutator defined on
homogeneous elements by:
\begin{equation}
\label{supercommutateur} [A,B]=A\circ B-(-1)^{p(A)p(B)}\,B\circ A.
\end{equation}
The space
$\cS\cP$ of superpseudodifferential symbols on $S^{1|1}$ has the
following form:
\begin{equation*} \cS\cP=C^{\infty}(S^{1|1})\otimes\Large(
\bbC[\xi][[ \xi^{-1}]] \oplus \bbC[\xi][[ \xi^{-1}]]\zeta\Large),
\end{equation*}
where $ \bbC[\xi][[ \xi^{-1}]]$ is the space of (formal)
Laurent series of finite order in $\xi.$

Any element of
$\cS\cP$ can be expressed in the following form:
\begin{equation*} S(x,\xi,\zeta)=
\sum_{-\infty}^n F_k(x)\xi^k+ \Big(\sum_{-\infty}^n
G_k(x)\xi^k\Big)\zeta,
\end{equation*}
where $F_k,G_k\in C^{\infty}(S^{1|1})$,
the symbol $\zeta=\overline{\theta}+\theta\xi$ corresponds to
$\eta$, $\xi$ corresponds to $\frac{\partial}{\partial x}$ and
$\overline{\theta}$ corresponds to $\frac{\partial}{\partial
\theta}$ (hence $ \overline{\theta}^2= \zeta^2=0$).

For $F\in C^{\infty}(S^{1|1})$, one has $\zeta
F\xi^m=\sigma(F)\xi^m\zeta$, so then, the multiplication in
$\cS\cP$ is obvious. On $\cS\cP$, there is a super Poisson bracket
given by (cf. \cite{GLS}):
\begin{equation}
\label{poissonbraket} \{S,T\}=\frac{\partial
S}{\partial\xi}\frac{\partial T}{\partial x}-\frac{\partial
S}{\partial x}\frac{\partial T}{\partial
\xi}-(-1)^{p(S)}\Big(\frac{\partial
S}{\partial\theta}\frac{\partial T}{\partial \overline{\theta} }
+\frac{\partial S}{\partial \overline{\theta} }\frac{\partial
T}{\partial\theta}\Big)\;\text{ for any $S,\; T\in\cS\cP$.}
\end{equation}
Consider a
family of associative laws on $\cS \Psi \cD \cO $ depending on one
parameter $h\in ]0,1]$ by:
\begin{equation*}
 F\,\eta\,^m \,\circ_h\, G\,\eta\,^n=
 \left\{
\begin{array}{ll}
 \displaystyle\sum_{k=0}^{\infty}\,(^m_k)_s\,
F\,h\,^{[\frac{k}{2}]}\,\eta\,^k(\sigma^{m-k}(G))\,\eta\,^{m+n-k}\;&\text{
if $m$ and $n$ are odd },\\[12pt]
\displaystyle\sum_{k=0}^{\infty}\,(^m_k)_s\,
F\,h\,^{[\frac{k-1}{2}]}\,\eta\,^k(\sigma^{m-k}(G))\,\eta\,^{m+n-k}\;&\hbox{
otherwise.}
\end{array}
\right.
\end{equation*}
Denote by $\cS \Psi \cD \cO_h$ the associative
superalgebra of superpseudodifferential operators on $S^{1|1}$
equipped with the multiplication $ \circ_h$. It is clear that all
the associative superalgebras $\cS \Psi \cD \cO_h $ are isomorphic
to each other.

For the supercommutator $[A,B]_h := \frac{1}{h}(A\circ_h
B-(-1)^{p(A)p(B)}B\circ_h A) $, one has:
\begin{equation*}
[A,B]_h=\{A,B\}+O(h),
\end{equation*}
and therefore
$\lim_{h\rightarrow0}[A,B]_h=\{A,B\},$ where we identify $\cS\cP$
with $\cS \Psi \cD \cO$ as vector spaces. Hence the Lie
superalgebra $\cS \Psi \cD \cO_{SL}$ contracts to the Poisson
superalgebra $\cS\cP$ (cf. \cite{EL}).

Furthermore, $\cS \Psi \cD \cO_{SL}$ admits
an analogue of the Adler trace defined on the Lie algebra $\Psi
\cD \cO$ of pseudodifferential operators on $S^1$ (cf. \cite{MR, AOR1}):
 let $A=\sum_{k\in\bbZ}F_{k}\eta^{k}$ be a superpseudodifferential
operator. Its super residue $\text{Sres}(A)$ is the coefficient
$F_{-1}\in C^{\infty}(S^{1|1})$ and the {\it Adler supertrace}
(which vanishes on the brackets)
is
\begin{equation}
\label{str} \text{Str}(A)=\int_{S^{1|1}}\text{Sres}(A)\;\vol(x,\theta)=
\int_{S^{1}}\frac{\partial F_{-1}}{\partial_{\theta}}dx.
\end{equation}

Recall that the Lie superalgebra $\cK(1)$ (also known as the
Neveu-Schwartz superalgebra without central charge, cf.
\cite{AOR1, AB}) consists of vector fields on $S^{1|1}$ preserving
the Pfaff equation given by the contact $1$-form $\alpha=dx+\theta
d\theta$. Explicitly
$\cK(1)$ consists of vector fields of the form:
\begin{equation*}
\label{contact} v_F = F\,\eta^2 +
\frac{1}{2}\,
\eta(F)\,\bar\eta,\qquad \hbox{where}~~
 \bar\eta = \frac{\partial}{\partial
\theta} - \theta \frac{\partial}{\partial x}\ .
\end{equation*}

\section{Statement of the problem}
The main purpose of this paper is to study  deformations of the
canonical embedding $\rho:{\cK(1)}\rightarrow{\cS \Psi \cD \cO}_{SL}$
defined by
\begin{equation}
\label{embedding}
\rho(v_F) = F\,\eta^2 +
\frac{1}{2}\,
\eta(F)\,\bar\eta
\end{equation}
into a one-parameter family of Lie superalgebra homomorphisms.
\subsection{ Formal deformations}
Let $\rho:\cK(1) \rightarrow{\cS \Psi \cD \cO}_{SL}$ be an embedding of Lie superalgebras,
\begin{equation}
\label{deformation}
\widetilde{\rho}_t=\rho+\sum_{k=1}^{\infty}t^{k}\rho_{k}~: \cK(1)
\rightarrow{\cS \Psi \cD \cO}_{SL},\quad \text{satisfying }\;
\widetilde{\rho}_{t}([X,Y]) =
[\widetilde{\rho}_t(X),\widetilde{\rho}_t(Y)],
\end{equation}
where  $\rho_k:\cK(1) \rightarrow{\cS \Psi \cD \cO}$ are even
linear maps, a {\it formal deformation} of $\rho$.

The bracket in the right hand side in (\ref{deformation}) is a natural
extension of the Lie bracket in ${\cS \Psi \cD\cO}_{SL}$ to ${\cS \Psi
\cD \cO}_{SL}[[t]].$ Two formal deformations ${\widetilde{\rho}}_t$ and
${\widetilde{\rho}}_t '$ are said to be {\it equivalent} if
there exists an inner automorphism $I_t:{\cS \Psi \cD
\cO}[[t]]\rightarrow {\cS \Psi \cD \cO}[[t]]$
\begin{equation}
\label{equivalent1} I_t=\exp(t \; \ad F_1 + t^2\; \ad F_2 +
\cdots),
\end{equation}
where $F_{i}\in {\cS \Psi \cD \cO}$ such
that $p(F_{i})=p(t^i)$, satisfying
\begin{equation}
\label{equivalent2} {\widetilde{\rho}}_t '
=I_t\circ{\widetilde{\rho}}_t.
\end{equation}
\subsection{ Polynomial deformations}
Observe
that a polynomial deformation defined in this section is NOT a particular case of a
formal definition. Recall that a deformation $\widetilde{\pi}$ of a homomorphism
$\pi:\Vect(S^1)\rightarrow \Psi\cD\cO$ defined by
\begin{equation*}
\pi(f(x)\,\partial_{x})=f(x)\,\xi
\end{equation*}
is (after \cite{OR1}) said to be {\bf polynomial}
if it is an homomorphism of the following
form
\begin{equation*}
 \widetilde{\pi}(c)= \pi + \sum_{k \in \bbZ }\,
\widetilde{\pi}_{k}(c)\,\xi^k,
\end{equation*}
where $c \in\bbR^n$ are parameters
of deformation, each linear map
$\widetilde{\pi}_{k}(c):\Vect(S^1)\rightarrow C^{\infty}(S^1)$
being  polynomial in $c$, $\widetilde{\pi}_k \equiv{0}$ for
sufficiently large $k$ and $\widetilde{\pi}_{k}(0)=0$.

Now, consider a Lie superalgebra
homomorphism $\widetilde{\rho}(c): \cK(1)\rightarrow\cS\Psi\cD\cO_{SL}$
of the following form:
\begin{equation}
\label{deformation poly} \widetilde{\rho}(c)= \rho + \sum_{k \in
\bbZ }\,\widetilde{\rho}_{k}(c),
\end{equation}
where $\widetilde{\rho}_{k}(c):\cK(1)\rightarrow \cS \cP_k $ are
even linear maps, polynomial in $c \in \bbR^{n}$ and such that
 $\widetilde{\rho}_k \equiv{0}$ for sufficiently large $k$ and
 $\widetilde{\rho}_{k}(0)=0$.

To define the notion of equivalence in the case of polynomial
deformations, one simply replaces the formal automorphism $I_t$ in
(\ref{equivalent1}) by an automorphism
\begin{equation}
\label{equivalent poly1} I(c):{\cS \Psi \cD \cO}_{SL}\longrightarrow{\cS
\Psi \cD \cO}_{SL}
\end{equation}
 depending on $c \in \bbR^n$ in the following way :
\begin{equation}
\label{equivalent poly2} I(c)=\exp( \sum_{i=1}^n c_{i}\; \ad F_{i}
+\sum_{i,j=1}^n c_{i}c_{j}\; \ad F_{i,j} + \cdots),
\end{equation}
where $ F_{i} , F_{i,j},\cdots F_{i_1\cdots i_k}$ are even
elements of ${\cS \Psi \cD \cO}$.

\begin{remark}
{\rm Theory of polynomial deformations seems to be richer than
that of formal ones. The equivalence problem for polynomial
deformations has additional interesting aspects related to
parameter transformations.}
\end{remark}
\section{Deformations and cohomology}
In this section, we will give a relationship between formal
and polynomial deformations of Lie superalgebra homomorphisms and
cohomology, cf. Nijenhuis and Richardson \cite{NR}.
\subsection{Infinitesimal deformations and the first cohomology}
If $\rho :\frak g \rightarrow \frak b$ is a Lie superalgebra
homomorphism, then $\frak b$ is naturally a $\frak g$-module.
A map $\rho + t\rho_1:\frak g \rightarrow \frak
b$, where $\rho_1 \in Z^1(\frak g,\frak b)$ is a Lie
superalgebra homomorphism up to quadratic terms in $t$, it is said
to be an {\bf infinitesimal deformation}.

The problem is now to find higher order prolongations of these
infinitesimal deformations. Setting $\varphi_t = \widetilde{\rho}_t - \rho$, one can
rewrite the relation (\ref{deformation}) in the following way:
\begin{equation}
\label{developping} [\varphi_t(X) , \rho(Y) ] + [\rho(X) ,
\varphi_t(Y) ] - \varphi_t([X , Y]) +\sum_{i,j > 0} \;[\rho_i(X) ,
\rho_j(Y)]t^{i+j} = 0\ .
\end{equation}
The first three terms are $(\delta\varphi_t) (X,Y)$, where
$\delta$ stands for the coboundary. For arbitrary linear maps $\varphi , \varphi' :
\frak g\longrightarrow\frak b$, define:
\begin{equation}
\label{maurrer cartan1}
\renewcommand{\arraystretch}{1.4}
\begin{array}{l}
{}[[\varphi , \varphi']] : \frak g \otimes \frak g
\longrightarrow\frak b\\
{}[[\varphi , \varphi']] (X , Y) =
[\varphi(X) , \varphi'(Y)] + [\varphi'(X) , \varphi(Y)].\end{array}
\end{equation}
The relation (\ref{developping}) becomes now equivalent to:
\begin{equation}
\label{maurrer cartan2} \delta\varphi_t + \frac{1}{2} [[\varphi_t
,  \varphi_t]] = 0.
\end{equation}
Expanding (\ref{maurrer cartan2}) in power series in $t$, we
obtain the following equation for $\rho_k$:
\begin{equation}
\label{maurrer cartan3} \delta\rho_k + \frac{1}{2} \sum_{i+j=k}
[[\rho_i ,  \rho_j]] = 0.
\end{equation}
The first nontrivial
relation is $\delta{\rho_2} + \frac{1}{2}  [[\rho_1 , \rho_1]] =
0$ gives the first obstruction to integration of an
infinitesimal deformation. Indeed, it is easy to check that for
any two  $1$-cocycles $\gamma_1$ and $\gamma_2 \in Z^1 (\frak
g , \frak b)$, the bilinear map $[[\gamma_1 , \gamma_2]]$ is a
$2$-cocycle. The first nontrivial relation (\ref{maurrer
cartan3}) is precisely the condition for this cocycle to be a
coboundary. Moreover, if one of the cocycles $\gamma_1$ or
$\gamma_2$ is a coboundary, then
$[[\gamma_1 , \gamma_2]]$ is a $2$-coboundary. We therefore,
naturally deduce that the  operation (\ref{maurrer cartan1})
defines a bilinear map:
\begin{equation}
\label{cup-product} H^1 (\frak g , \frak b)\otimes H^1 (\frak
g , \frak b)\longrightarrow H^2 (\frak g , \frak b),
\end{equation}
called the {\it cup-product}.

All the obstructions lie in $H^2 (\frak g , \frak b)$ and they
are in the image of $H^1 (\frak g , \frak b)$ under the
cup-product. So, in our case, we have to compute $H^1(\cK(1),\cS
\Psi \cD \cO)$ and the product classes in $H^2 (\cK(1),\cS \Psi
\cD \cO)$.
\section{The space $H^1(\cK(1),\cS \Psi \cD \cO$)}
\subsection{A filtration on $\cS \Psi \cD \cO$}
The natural embedding of $\cK(1)$ into $\cS \Psi \cD \cO$ given by
the expression (\ref{embedding}) induces a $\cK(1)$-module
structure on $\cS \Psi \cD \cO$. Analogously, we have a
$\cK(1)$-module structure on $\cS\cP$ given by the natural
embedding of
 $\cK(1)$:
\begin{equation}
\label{embedding0} \overline\pi~:~
v_F\mapsto F\,\xi +
\frac{1}{2}\,\eta(F)\,\bar\zeta,
\end{equation}
where $\bar\zeta = \overline{\theta} - \theta\xi$.

Setting the degree of $x, \theta$ be zero and the degree
of $\xi, \zeta$ be $1$ we introduce a $\bbZ$-grading in the
Poisson superalgebra $\cS\cP$. Then we have
\begin{equation}
\label{grading} \cS\cP=\widetilde{\bigoplus}_{n\in \bbZ}S\cP_n
:=(\bigoplus_{n<0}S\cP_n) \bigoplus(\prod_{n\geq 0}S\cP_n),
\end{equation}
\noindent where $S\cP_n=\{F\xi^{-n}+G\xi^{-n-1}\zeta\mid F,\, G\in
C^{\infty}(S^{1|1})\}$ is the homogeneous subspace of degree $-n$.
\bigskip
Each element of $\cS\Psi \cD \cO$  can be expressed as
 \begin{equation*}
A=\sum_{k\in \bbZ}(F_k+G_k\eta^{-1})\eta^{2k}, ~~ \mbox{where}
~~F_k,\, G_k\in C^{\infty}(S^{1|1}).
\end{equation*}
We define the
order of $A$ to be
 \begin{equation*}
\text{ord}(A)=\sup\{k~\mid~ F_k\neq 0 \hbox{ or }
G_k\neq 0\}.
\end{equation*}
This definition of order equips  $\cS\Psi \cD \cO$
with a decreasing filtration as follows: \noindent set
\begin{equation*}
{\bf F}_n=\{A\in \cS\Psi \cD \cO~\mid~\text{ord}(A)\leq -n\},~~ \mbox{where}~
n\in \bbZ.
\end{equation*}
So one has
\begin{equation}
\label{filtration} \ldots\subset {\bf F}_{n+1}\subset {\bf
F}_n\subset \ldots.
\end{equation}
This filtration is compatible with the multiplication and the super Poisson
bracket, that is, for $A\in {\bf F}_n$ and $B\in {\bf F}_m$, one
has $A\circ B\in {\bf F}_{n+m}$ and $\{A,B\}\in {\bf F}_{n+m-1}$,
after we identify $\cS\cP$ with $\cS\Psi \cD \cO$. This filtration
makes $\cS\Psi \cD \cO$ an associative filtered superalgebra.
Moreover, this filtration is compatible with the natural action of
$\cK(1)$ on $\cS\Psi \cD \cO$. Indeed, if $v_F\in \cK(1)$ and
$A\in {\bf F}_n$, then
\begin{equation*}
v_F\cdot A=[v_F,A]\in {\bf F}_n.
\end{equation*}
The induced $\cK(1)$-action on the quotient ${\bf F}_n/{\bf F}_{n+1}$
is isomorphic to the $\cK(1)$-action on $\cS\cP_n$.
 Therefore, the $\cK(1)$-action on the associated
graded space of the filtration (\ref{filtration}), is isomorphic
to the graded $\cK(1)$-module $\cS\cP$, that is
\begin{equation*}
S\cP\simeq \widetilde{\bigoplus}_{n\in  \bbZ}{\bf F}_n/{\bf F}_{n+1}.
\end{equation*}

Now we can deduce the cohomology of the filtered module from the
cohomology of the associated graded module.
\subsection{$H^1(\cK(1),\cS\cP)$}
Observe that $H^1(\cK(1),\cS\cP)=\bigoplus_{n\in \bbZ}H^1(\cK(1),\cS\cP_n)$.
These spaces are known (see \cite{AB}). They are
nontrivial if and only if $n=0, 1$ and the corresponding
dimensions are $3$ and $1$, respectively.
Therefore, $ H^1(\cK(1),\cS\cP)\cong\bbR^4$. The nontrivial
cocycles generating the space $H^1(\cK(1),\cS\cP)$ are
($\ad_{\zeta}(\overline\pi(v_F))=\{\zeta,\overline\pi(v_F)\}$ with
$\overline\pi$ as in (\ref{embedding0})):
\begin{gather}
C_0(v_F) =\displaystyle
\frac{1}{4}(F+\sigma(F))+\half F, \\
C_1(v_F)
=\displaystyle \eta^2(F),\label{ga1} \\
C_{2}(v_F)=\displaystyle
\ad^3_{\zeta}(\overline\pi (v_F))\xi^{-2}\overline{\zeta}\
\end{gather}
with values in $\cS\cP_0$, and
\begin{equation}
\label{ga2}
 \displaystyle C_{3}(v_F)= \displaystyle
\ad^5_{\zeta}(\overline\pi(v_F))\xi^{-3}\overline{\zeta}
\end{equation}
with values in $\cS\cP_1$.
\subsection{$H^1(\cK(1),\cS \Psi \cD \cO )$ (\cite{AB})}
The result of \cite{AB} is a specialization at $h=1$ of the following theorem
 obtained as in \cite{AB}:
\begin{theorem}
\label{th2} The space $H^1(\cK(1),\cS \Psi \cD\cO_h )$ is purely
even. It is spanned by the classes of the following nontrivial
$1$-cocycles
$$
\renewcommand{\arraystretch}{1.4}
\begin{array}{ll}
\label{excocycles3} \Theta_{0}(v_F)=&\displaystyle
\frac{1}{4}(F+\sigma(F))+\half F, \\
\Theta_{1}(v_F) =&\displaystyle \eta^2(F), \\
\Theta_{2_h}(v_F)= &\displaystyle \sum_{n=1}^{\infty}(-1)^n
h^{n-1}\, \frac{n-2}{n}\sigma(\bar{\eta}\,^{2n+1}(F))
\bar{\eta}\,^{-2n+1}+\\
&\sum_{n=1}^{\infty}(-1)^n\,h^n\,
\frac{n-3}{n+1}\bar{\eta}\,^{ 2n+2}(F) \bar{\eta}\,^{-2n},\\
\displaystyle
\Theta_{3_h}(v_F)=&\displaystyle\sum_{n=2}^{\infty}(-1)^n\,
h^{n-2}\, \frac{n-1}{n}\sigma(\bar{\eta}\,^{2n+1}(F))
\bar{\eta}\,^ {-2n+1}+\\
&\sum_{n=2}^{\infty}(-1)^n h^{n-1}\,
\frac{n-1}{n+1}\bar{\eta}\,^{2n+2}(F) \bar{\eta}\,^{-2n}\ .
\end{array}
$$
\end{theorem}

\section{Integrability of infinitesimal deformations}
The space $H^1 (\cK(1) , \cS \Psi \cD \cO )$
classifies infinitesimal deformations of the standard embedding
$\cK(1)\longrightarrow{\cS \Psi \cD \cO}_{SL}$ given by (\ref{embedding}). In
this section we will calculate the integrability conditions of
infinitesimal deformations into polynomial ones.
Any nontrivial infinitesimal deformation can be expressed in the
following form:
\begin{equation}
\rho_1=\rho+\sum_{0\leq i\leq 3}{}c_i\,\Theta_i\;,\text{ where $c_0,~c_1,~c_2,~c_3\in\bbR$.}
\end{equation}
The integrability condition (below) imply that either $c_0=0$ or $c_2=c_3=0$. 

\subsection{Deformations generated by $\Theta_0$ and $\Theta_1$}
Since zero-order operators commute in ${\cS \Psi \cD \cO}$, it is
evident that the cup-products $ [[\Theta_0, \Theta_0]]$,
$[[\Theta_0, \Theta_1]]$ and $[[\Theta_1, \Theta_1]]$ vanish
identically, and therefore the map
\begin{equation}
\label{deformation inf}
\rho_{\nu,\lambda}:\cK(1)\rightarrow\cS
\Psi \cD \cO,~v_F\mapsto\rho_{\nu,\lambda}\,(v_F) = \rho\,(v_F) +
\nu\;\Theta_0(v_F) + \lambda\,\Theta_1(v_F)
\end{equation}
is indeed, a nontrivial deformation of the standard embedding.
This deformation is polynomial since it is of order 1.
\begin{proposition}\label{ord1} Any nontrivial formal deformation of the embedding
(\ref{embedding}) generated by $\Theta_0$ and $\Theta_1$ is
equivalent to a deformation of order 1, that is, to a deformation
given by (\ref{deformation inf}).
\end{proposition}
\begin{proof}
Consider a  formal deformation of  the embedding (\ref{embedding})
generated by $\Theta_0$ and $\Theta_1$:
\begin{equation}
\label{deformation f} \widetilde{\rho}_t=\rho+t_0\,\Theta_0 +
t_1\,\Theta_1+ \sum_{m\geq2}\,\sum_{
i+j=m}{}\,t_0^i\,t_1^j\,\rho\,^{(m)}_{i,j},
\end{equation}
where the highest-order terms $\rho^{(m)}_{ij}$ are even linear
maps from ${\cK(1)}$ to ${\cS \Psi \cD \cO}$.
The solution $\rho^{(2)}_{ij}$ of (\ref{maurrer
cartan3}) is defined up to a 1-cocycle and it has been shown
in~\cite{aalo2, ff2} that different choices of solutions of
(\ref{maurrer cartan3}) correspond to equivalent deformations.
Thus, one can always kill $\rho^{(2)}_{ij}$. Then, by recurrence,
the highest-order terms satisfy the equation $\d\rho^{(m)}_{ij}=0$
and can also be killed.
\end{proof}
\begin{remark}
{\rm Recall that, in the classical case (cf.
\cite{OR1}), there exists an analogous deformation
$\pi_{\nu,\lambda}$ of $\pi$. Then, one can easily check that the following diagram
commutes: \vspace{0,4cm}

 \hspace{0.4cm} $\cK(1)\stackrel{\rho_{\nu,\lambda}}\longrightarrow
 \cS \Psi \cD \cO$

\hspace{0.5cm} $\uparrow i \hspace{1.3cm} \uparrow j$

\hspace{0.4cm}
$\Vect(S^1)\stackrel{\pi_{\nu,\lambda}}\longrightarrow
\Psi\cD\cO$\vspace{0,3 cm}\\
where
\begin{gather*}
 i(f(x)\partial_x)=v_{f(x)},\\
\pi_{\nu,\lambda}(f(x)\partial_x)= f(x)\xi+\nu f(x)+\lambda
f'(x), \\
j(A)=A+\frac{1}{2}\eta^2(\partial_{\xi}A)\theta\eta\ .
\end{gather*}
}
\end{remark}

Real difficulties begin
when we deal with polynomial or formal integrability of the infinitesimal deformations
corresponding to the cocycles $\Theta_1, \Theta_2$ and $\Theta_3$.
\subsection{Deformations generated by $\Theta_1$, $\Theta_2$ and $\Theta_3$}
Consider an infinitesimal deformation of the standard embedding of
$\cK(1)$ into ${\cS \Psi \cD \cO}_{SL}$ defined by the cocycles
$\Theta_1,\Theta_2,\Theta_3$ and depending on the real
parameters $c_1, c_2, c_3$
\begin{equation}
\label{integrability} \widetilde{\rho}(c)(v_F)=\rho (v_F) +
c_1\Theta_1(v_F) + c_2\Theta_2(v_F) + c_3\Theta_3(v_F).
\end{equation}
\begin{theorem}
\label{Main} The infinitesimal deformation (\ref{integrability})
corresponds to a polynomial deformation, if and only if the
following relations are satisfied :
\begin{equation}
\label{condition necessary1} \left\{
\begin{array}{ll}
3c_1c_3 - 2c_1^3 -2c_1^{2}c_3
+ c_1^2 + 2c_3^2 = 0\\
c_1 = c_2
\end{array}
\right.\\
\end{equation}
or
\begin{equation}
\label{condition necessary2} \left\{
 \begin{array}{ll}
 c_3c_1 - 2c_3c_1^2 -2c_3^2 = 0\\
 c_2 = 0
 \end{array}
 \right.
\end{equation}
\end{theorem}
\noindent To prove Theorem \ref{Main}, we first introduce the
notion of homogeneity for a deformation given by differentiable
maps, then we will prove that the conditions (\ref{condition
necessary1}--\ref{condition necessary2}) are necessary for
integrability of infinitesimal deformations. In the end of this
section we will show that these relations are sufficient by
exhibiting explicit deformations.
\subsubsection{Homogeneous deformation}
Consider an arbitrary {\bf polynomial} deformation of the standard
embedding, corresponding to the infinitesimal deformation
(\ref{integrability}):
\begin{gather}
 \widetilde{\rho}(c)(v_{F})= \rho(v_F)+
 c_1  \eta^2(F) + c_2 (\sigma(\bar\eta^3(F))\bar\eta^{-1} +
 \bar\eta^4(F)\bar\eta^{-2})\notag\\
 ~~\qquad\qquad{}+ c_3 (\sigma(\bar\eta^5(F)\bar\eta^{-3}) +
 \sum_{k\in\bbZ}P_k(c)\rho_k(v_F)\ ,\label{polynomial}
  \end{gather}
 where $c = (c_1, c_2, c_3)\in\bbR^3$, $ P_k$ are polynomial functions of degree $\geq
 2$ and $\rho_k$ are some differentiable
even linear  maps from $\cK(1)$ to $\cS\cP_k$.

Note that, since the cocycles $\Theta_1, \Theta_2, \Theta_3$
 are defined by differentiable maps, an arbitrary solution of the
 deformation problem is also defined via differentiable maps. This
 follows from the Gelfand-Fuchs formalism of differentiable (or
 local) cohomology (see \cite{Fu}).

Now, let us introduce a notion of homogeneity for deformation
given by differentiable maps.
A deformation (\ref{polynomial}) is said to be
{\it homogeneous of degree $m$} if $\widetilde{\rho}(c)(v_{F})$
is of the form:
\begin{equation*}
  \widetilde{\rho}(c)(v_{F})=
\sum_{k\in\bbZ}P_k(c)(\sigma^k(\bar\eta^{k+m}(F)))\bar\eta^{-k}.
\end{equation*}
Since the cocycles $\Theta_1, \Theta_2$ and $ \Theta_3$ are
of degree $2$, every homogeneous deformation
(\ref{polynomial})
 corresponding to a nontrivial infinitesimal deformation is of
 degree $2$:
 \begin{gather}
 \widetilde{\rho}(c)(v_{F})= \rho(v_F) +
 c_1  \eta^2(F) + c_2 (\sigma(\bar\eta^3(F))\bar\eta^{-1} +
 \bar\eta^4(F)\bar\eta^{-2})\notag\\
 ~~\qquad\qquad{}+ c_3 (\sigma(\bar\eta^5(F)\bar\eta^{-3}) +
 \sum_{k\geq 4}P_k(c)(\sigma^k(\bar\eta^{k+2}(F)))\bar\eta^{-k}.
 \label{def homogene}
 \end{gather}
\begin{proposition}
\label{ho} Every deformation (\ref{polynomial}) is
equivalent to a homogeneous deformation (\ref{def homogene}).
\end{proposition}
\begin{proof}
It is easy to see that any homomorphism preserves homogeneity. This means that
the first term in (\ref{polynomial}) (the term of the lowest
degree in $c$) which is not homogeneous of degree $2$ must lie in
$H^1(\cK(1),\cS\cP)$. 
Such a 1-cocycle is cohomologous to a linear combination of the
1-cocycles $C_1, C_2$ and $C_3$, see (\ref{ga1})--(\ref{ga2})
which are homogeneous of degree 2. Thus, one can add
(or remove) a coboundary in the term of the polynomial deformation
(\ref{polynomial}) to obtain an equivalent one.
\end{proof}

\subsubsection{Integrability conditions are necessary}
The infinitesimal deformation (\ref{integrability})
 is clearly of the form
\begin{gather}
\widetilde{\rho}(c)(v_{F})= \rho(v_F) +
 c_1  \eta^2(F)\notag \\
 ~~\qquad\qquad{}+ c_2 (\sigma(\bar\eta^3(F))\bar\eta^{-1} +
 \bar\eta^4(F)\bar\eta^{-2})\notag\\
 ~~\qquad\qquad{}+ c_3 \sigma(\bar\eta^5(F))\bar\eta^{-3} + \cdots\ ,
 \label{debut}
\end{gather}
where $\lq\lq\cdots"$ means the terms in $\bar\eta^{-4},
\bar\eta^{-5}$. To compute the obstructions for integrability of
the infinitesimal deformation (\ref{integrability}), one has to
add the first nontrivial terms and impose the homomorphism
condition. So, put
\begin{equation}
\label{second def}
\bar{\rho}(c)(v_{F})=\widetilde{\rho}(c)(v_{F})+
P_{4}(c)\bar{\eta}^{6}(F) \bar{\eta}^{-4}+
P_{5}(c)\sigma({\bar{\eta}}^{7}(F)) \bar{\eta}^{-5},
\end{equation}
where $P_{4}(c)$ and $ P_{5}(c)$ are some polynomials in
$c=(c_{1},c_{2},c_{3})$ and compute the difference
\begin{equation*}
[\bar\rho(c)(v_F) , \bar\rho(c)(v_G)] -
\bar\rho(c)([v_{F},v_{G}]).
\end{equation*}
A straightforward but boring
computation leads to the following equations:
\begin{equation}
\label{n_1}
\renewcommand{\arraystretch}{1.4}
\begin{array}{l}
c_2c_1 = c_2^2\,,\\
 3P_4 = 2c_3 - c_2 - 2c_3c_1 + 4c_3c_2 + c_2c_1\,,\\
 3P_5 = -c_2 - 4c_3 + 2c_2^2 - 2c_3c_2 + 4c_3c_1.\end{array}
\end{equation}
Let us go one step further, expand our deformation up to
$\bar\eta^{-7}$, that is, put
\begin{equation*}\bar{\bar{\rho}}(c)(v_{F})=\bar{\rho}(c)(v_{F})+
P_{6}(c)\bar{\eta}^{8}(F) \bar{\eta}^{-6}+
P_{7}(c)\sigma({\bar{\eta}}^{9}(F)) \bar{\eta}^{-7}.
\end{equation*}
The homomorphism condition leads to a nontrivial relations for the
parameters. For $c_1 = c_2$, the relations are:
\begin{equation}
\label{omr6}
\renewcommand{\arraystretch}{1.4}
\begin{array}{l}
2P_6 = -c_1 + c_1^2 + P_4(-3 + 2c_1)\,,\\
2P_7 = 5c_3 - 2c_3^2 - 6c_3c_1 - 3P_4 + 4c_1P_4\,,\\
\end{array}
\end{equation}
\begin{equation}
\label{omr67}
5P_7 = c_1 - 2c_1^2 + c_1P_4 + (3c_1 -
\frac{9}{2})P_5 - \frac{3}{2}P_6.
\end{equation}
Substituting expressions (\ref{n_1}), 
and (\ref{omr6}) for $P_4, P_5, P_6$
and $P_7$ in (\ref{omr67}), one gets formula
(\ref{condition necessary1}).

For $c_2 = 0$, the relations are:
\begin{equation}
\label{nizar6}
\renewcommand{\arraystretch}{1.4}
\begin{array}{l}
P_6 = (\frac{3}{2} - c_1)(P_4 + P_5)\,,\\
2P_7 =3c_3 - 4c_3c_1 - 2c_3^2\,,\\
\end{array}
\end{equation}
\begin{equation}\label{nizar67}
4(P_6 + P_7) = (-3 + 2c_1)(P_4 + P_5).
\end{equation}
Substituting expressions (\ref{n_1}), 
and
(\ref{nizar6}) for $P_4, P_5, P_6$ and $P_7$ in
(\ref{nizar67}), we get formula (\ref{condition
necessary2}). We have thus shown that the conditions
(\ref{condition necessary1})
are
necessary for integrability of infinitesimal deformations.
\begin{remark}
{\rm The obstructions to
integrability of an infinitesimal deformation
(\ref{integrability}) which does not satisfy the conditions
(\ref{condition necessary1})
corresponds to a nontrivial class of $H^2(\cK(1), \cS\cP_3)$. }
\end{remark}
\subsection{Introducing the parameter h}
One can now modify the relations in order to get a deformation in
$\cS\Psi\cD\cO_h$, the scalar $h$ then appears with different
powers according to the \lq\lq weight" of the respective terms in
 formulas (\ref{condition necessary1}--\ref{condition
necessary2}). One finally gets
\begin{equation}
\label{condition necessary h1} \left\{
\begin{array}{ll}
h^{2}c_1^2+ h(3c_1c_3 - 2c_1^3) + 2c_3^2 -2c_1^{2}c_3 = 0\\
c_1 = c_2
\end{array}
\right. \quad
\text{ or }\quad
 \left\{
 \begin{array}{ll}
 hc_3c_1 - 2c_3c_1^2 -2c_3^2 = 0\\
 c_2 = 0 \cdot
 \end{array}
 \right.
\end{equation}
These relations are necessary for the integrability of the
infinitesimal deformation (\ref{integrability}) in
$\cS\Psi\cD\cO_h$.
\begin{remark}
{\rm  By setting weights: $wht(c_1)=wht(h)=1$ and $wht(c_3)=2$,
we make the polynomials (\ref{condition necessary
h1}) 
homogeneous of weight 4.
Moreover, setting $h=0$, one gets the necessary conditions to
have a polynomial deformation of the standard embedding
(\ref{embedding0}) corresponding to a given infinitesimal one
generated by the cocycles $C_1, C_2$ and $C_3$ given by (\ref{ga1})--(\ref{ga2}).}
\end{remark}

Now, we will give a natural description of the curves defined by
equations (\ref{condition necessary h1}) 
in order to unveil their algebraic nature.
\subsubsection{A rational parameterization}
There exists a rational parameterization of the curves
(\ref{condition necessary h1}): 
\begin{proposition}  ~~i) For all $\lambda \in\bbR$, the constants
\begin{equation}
\label{parameter1} \left\{
\begin{array}{lll}
 c_1 &=& -\lambda \\
c_2 &=& -\lambda \\
c_3 &=& h\lambda
\end{array}
\right.\quad
\text{ or }\quad
\left\{
\begin{array}{lll}
 c_1 &=& -\lambda \\
c_2 &=& -\lambda \\
c_3 &=& \lambda^2 + \half h\lambda
\end{array}
\right.
\end{equation}
satisfy the first of relations (\ref{condition necessary h1}).
\medskip

ii) For all $\lambda \in\bbR$, the constants
\begin{equation}
\label{parameter3} \left\{
\begin{array}{lll}
 c_1 &=& -\lambda \\
c_2 &=& c_3 = 0
\end{array}
\right.
\quad
\text{ or }\quad
\left\{
\begin{array}{lll}
 c_1 &=& -\lambda \\
c_2 &=& 0\\
c_3 &=& -\lambda^2 - \half h\lambda
\end{array}
\right.
\end{equation}
satisfy the second of relations (\ref{condition necessary h1}).
\medskip

iii) Any triple $c_1, c_2, c_3 \in \bbR$ satisfying
(\ref{condition necessary h1}) 
is of the form (\ref{parameter1}) 
 or(\ref{parameter3}) 
for same $\lambda$. 
\end{proposition}
\begin{proof}
By direct computations.
\end{proof}
\begin{remark}
{\rm Geometrically, the curves (\ref{condition necessary1}) and
(\ref{condition necessary2}) are just lines and parabolas.  }
\end{remark}
\noindent Now, the analogue of Richardson-Nijenhuis theory in
supergeometry (cf. (\ref{cup-product})) prescribes us to compute
$H^2(\cK(1),\cS\Psi\cD\cO)$ in order to obtain the complete
information concerning the cohomological obstructions. This,
however, seems to be a quite difficult problem. We shall not do
that; an explicit construction of deformations will allow to avoid
the standard obstruction framework.
\subsubsection{Construction of deformations}
Now, to complete the proof of our main result (Theorem
\ref{Main}), we construct a polynomial deformation corresponding
to any infinitesimal deformation (\ref{integrability}) satisfying
the condition (\ref{condition necessary h1}). 
This implies that these conditions are not only
necessary, but also sufficient for integrability of infinitesimal
deformations (\ref{integrability}).

The space $H_0^1(\cS\Psi\cD\cO,\cS\Psi\cD\cO)$ of even outer
superderivations of the Lie superalgebra $\cS\Psi\cD\cO$ contains
the linear operator $\ad\log\xi$ on $\cS\Psi\cD\cO$ (cf.
\cite{AOR1}). This outer superderivation can be integrated to a
one-parameter family of outer automorphisms denoted by $\Psi_\nu$
and defined by
\begin{equation}
\label{auto} \Psi_\nu(F) = \xi^\nu \circ F \circ \xi^{-\nu},
\end{equation}
which should be understood as a Laurent series in $\overline\eta$
(depending on the parameter $\nu$).

Let us apply the
automorphism (\ref{auto}) to the elementary deformation
$\rho_{0,\lambda}$ (\ref{deformation inf}):
\begin{gather}
\widetilde\rho_1^{\lambda}(v_F)=
\Psi_{\frac{-2\lambda}{h}}(\rho(v_F) +
\lambda\eta^2(F))\notag\\
~~~\qquad{} = \rho(v_F) - \lambda\eta^2(F)\notag\\
\qquad\qquad{}- \lambda(\sigma(\bar\eta^3(F))\bar\eta^{-1} +
 \bar\eta^4(F)\bar\eta^{-2})\notag\\
 \qquad\qquad{}+ (\lambda^2 + \half \lambda h)(\sigma(\bar\eta^5(F)\bar\eta^{-3})
 +\cdots\
 \label{map1}
 \end{gather}
Since $\Psi_{\frac{-2\lambda}{h}}$ is an automorphism, it is,
indeed, a polynomial deformation of embedding (\ref{embedding})
for any $\lambda \in \bbR$, corresponding to any infinitesimal
deformation (\ref{integrability}) satisfying the second of
conditions (\ref{parameter1}).
\begin{proposition}
\label{Comain} The map
\begin{equation}
\label{explicite def} \widetilde\rho_{2}^{\lambda}: v_{F}
\rightarrow \rho(v_{F}) + \lambda\; \widetilde\Theta_{h}(v_{F}),
\end{equation}
where
$\widetilde\Theta_{h}=2h\Theta_{3_h}-\Theta_{2_h}-\Theta_{1}$, is
both a polynomial and a formal deformation of embedding
(\ref{embedding}) for any $\lambda \in \bbR$, corresponding to any
infinitesimal deformation (\ref{integrability}) satisfying the
first of conditions (\ref{parameter1}).
\end{proposition}
\begin{proof}
Since $\widetilde\Theta_{h}$ is an even $1$-cocycle, the map
$\widetilde\rho_2^{\lambda}$ is a polynomial deformation if the
supercommutator $[\widetilde\Theta_{h},~\widetilde\Theta_{h}]$
vanishes. So put
\begin{gather*}
\widetilde\Theta_{h}(v_{F})=\sum_{n\geqslant1}(-1)^n h^{n-1}
a^{(n+1)}\xi^{-n} \theta\frac{\partial}{\partial \theta} +
\sum_{n\geqslant0}(-1)^{n+1} h^{n} a^{(n+1)}\xi^{-n}\\
 ~~\qquad\qquad{}+2\sum_{n\geqslant1}(-1)^{n} h^{n-1} b^{(n)}\xi^{-n}
\frac{\partial}{\partial \theta}\ ,
\end{gather*}
where $F = a + 2b\theta$~ with~ $a,\;b\in C^{\infty}(S^1)$~
and compute
$[\widetilde\Theta_{h}(v_F),~\widetilde\Theta_{h}(v_G)]$, where
$G = c+2d\theta$ with $c,\;d \in C^{\infty}(S^1)$. Collect the
terms with $a\;^{(\alpha
+1)}\;\;c^{(\beta + 1)}\;\xi^{- \alpha - \beta}$ for $\alpha,\;\beta\;\in\;\bbN$:
\begin{equation}
\label{coeficient1} \left\{
\begin{array}{ll}
H(\alpha,\beta)~&
\mbox{if}\;\alpha\geq 2 \;\mbox{ and }~\beta\geq 2,\\
0~&\mbox{ otherwise },
\end{array}
\right.
\end{equation}
where
\begin{equation}
$$
\renewcommand{\arraystretch}{1.4}
\begin{array}{l}
H(\alpha,\beta) = -(-h)\,^{\alpha+\beta-1}\, \times
\left(\binom{\alpha + \beta - 1}{\alpha - 1}
 - \binom{\alpha + \beta - 1}{
 \beta - 1}
 + \mathop{\sum}\limits_{ n=1}^{ \beta-1}
\binom{\alpha + n - 1 }{ \alpha - 1}
 - \mathop{\sum}\limits_{ k=1} ^{ \alpha
-1}\binom{\beta + k - 1}{
 \beta - 1}
\right)\label{cocoeficient1}.
\end{array}
$$
\end{equation}
It is now easy to check that the expression (\ref{cocoeficient1})
vanishes for $\b=2$ and $\alpha\geq2$. We will prove by recurrence
that this expression vanishes for $\alpha\geq2$ and $\beta\geq2$.

Assume that, for $\beta\geq2$, one has
\begin{equation}
\label{recurrence}
\binom{\alpha + \beta - 1}{\beta -1} -
\binom{\alpha + \beta -1}{\alpha - 1}
 - \sum_{n=1}^{\beta
-1}\binom{\alpha + n -1}{\alpha -1}
 + \sum_{n=1}^{\alpha
-1}\binom{\beta + n -1}{ \beta - 1}
=0.
\end{equation}
Using that
\begin{gather*}
  \sum_{k=1}^{\alpha}\binom{\beta + k - 1}{\beta}
 =\binom{\alpha + \beta}{\alpha - 1}, \qquad
\binom{\alpha + \beta }{\beta}
 =\displaystyle \frac{\alpha + \beta}{\beta}\;
 \binom{\alpha + \beta -1 }{ \beta - 1}, \\
\binom{\alpha + \beta }{ \alpha - 1}
= \frac{\alpha + \beta}{\beta + 1}\;
\binom{ \alpha + \beta -1}{\alpha - 1},
\end{gather*}
and equation (\ref{recurrence}), one obtains
\begin{equation*}
\binom{\alpha + \beta }{\alpha}
-\binom{\alpha + \beta }{\alpha - 1}
-\sum_{n=1}^{\beta }
\binom{ \alpha + n -1 }{ \alpha - 1}
+\sum_{n=1}^{\alpha -1}
\binom{ \beta + n}{ \beta}
=0
\end{equation*}
which implies that the
expression (\ref{coeficient1}) vanishes.

Note that the term with
$a\,^{(\alpha + 1)}\,c\,^{(\beta + 1)}\,\xi\,^{-\alpha - \beta}
\theta\,\frac{\partial}{\partial \theta}$, where
$\alpha,\,\beta\,\in\,\bbN$, vanishes since it has the same expression
as (\ref{coeficient1}). Finally, one can easily see that the
coefficients of  $a\,^{(\alpha
+1)}\,d\,^{(\beta)}\,\xi^{-\alpha-\beta}\,\partial_{\theta}$ and
of
$c\,^{(\alpha +1)}\,b\,^{(\beta)}\,\xi\,^{-\alpha-\beta}\,\theta $ are the
same, and hence
\begin{equation}
\label{coeficient2} \left\{
\begin{array}{ll}
L(\alpha, \beta)~& \mbox{if}~\alpha\geq 1~\mbox{and}~\beta\geq2,\\[12pt]
0~&\mbox{otherwise},
\end{array}
\right.
\end{equation}
where
\begin{equation}
\label{cocoeficient2} L(\alpha , \beta) = -(-h)\,^{\alpha+\beta-2}
 \left( 1 -
\binom{\alpha + \beta -1 }{ \beta - 1}
 + \sum_{n=1}^ {\beta -1}
\binom{\alpha + n -1 }{ \alpha - 1}
 \right).
\end{equation}
The same arguments show that expression (\ref{cocoeficient2})
vanishes. We have proved thus that
$[\widetilde\Theta_{h},\widetilde\Theta_{h}]=0$. Hence, the map
$\widetilde\rho_2^{\lambda}$
 is both a polynomial and formal deformation. To
 complete the proof of Proposition (\ref{Comain}), observe
  that every infinitesimal deformation
(\ref{integrability}) satisfying the first of  conditions
(\ref{parameter1}) can be realized as the infinitesimal part of
the polynomial deformation $\widetilde{\rho}_2^{\lambda}$.
\end{proof}

Finally, we will construct a polynomial deformation corresponding
to any infinitesimal deformation (\ref{integrability}) satisfying
the second of conditions (\ref{parameter3}). We apply the
automorphism (\ref{auto}) to the polynomial deformation
(\ref{explicite def}):
\begin{gather}
\widetilde\rho_3^\lambda(v_F) =
\Psi_{\frac{-2\lambda}{h}}\circ\widetilde\rho_2^{-\lambda}(v_F)
= \rho(v_F) - \lambda\eta^2(F)
- (\lambda^2 +\half \lambda
h)(\sigma(\bar\eta^5(F)\bar\eta^{-3})
 +\cdots,\label{map2}
\end{gather}
so we obtain a polynomial deformation corresponding to any
infinitesimal deformation (\ref{integrability}) satisfying the
second of  conditions (\ref{parameter3}).
\begin{remark}
{\rm Under the first of conditions (\ref{parameter3}), any
infinitesimal deformation (\ref{integrability}) becomes a
polynomial deformation.}
\end{remark}

Applying the contraction procedure as $h\rightarrow 0$ to the
deformations (\ref{map1}), (\ref{explicite def}) and (\ref{map2}),
we get polynomial deformations of the infinitesimal deformation of
the standard embedding (\ref{embedding0}) generated by the
1-cocycles $C_1, C_2$ and $C_3$ corresponding to the conditions
(\ref{condition necessary h1}) 
at
$h=0$. More precisely, we get:
\begin{theorem}
Every nontrivial polynomial deformation of the standard embedding
(\ref{embedding0}) is equivalent to one of the four following
deformations: \begin{equation*}
\begin{array}{lll}
\rho_{1}^{\lambda}(v_{F})&=&\frac{1}{2}\,\big((F+\sigma(F))\,\xi+\eta(F)\,\zeta
\;\big)+ \lambda\,\eta^{2}(F),\\[12pt]
\rho_{2}^{\lambda}(v_{F})
&=&\frac{1}{2}\,\big((F+\sigma(F))\,\xi + \eta(F)\,\zeta \;\big)+ \lambda
\,\big(\eta^{2}(F) - \sigma(\bar{\eta}^{3}(F))\,\xi^{-1}\,\bar{\zeta}\;
\big),\\[12pt]
\rho_{3}^{\lambda}(v_{F})&=&
\frac{1}{2}\,\big((F(x-\frac{2\lambda}{\xi})+
\sigma(F(x-\frac{2\lambda}{\xi})))\,\xi + \eta(F(x-\frac{2\lambda}{\xi}))\,\zeta\;\big)
+ \lambda\,\eta^{2}(F(x-\frac{2\lambda}{\xi})), \\[12pt]
\rho_{4}^{\lambda}(v_{F})&=&
\frac{1}{2}\,\big((F(x-\frac{2\lambda}{\xi})+
\sigma(F(x-\frac{2\lambda}{\xi})))\,\xi + \eta(F(x-\frac{2\lambda}{\xi}))\,\zeta
\;\big) + \lambda\,\big(\eta^{2}(F(x-\frac{2\lambda}{\xi}) \\[12pt]
& &
-\sigma(\bar{\eta}^3(F(x-\frac{2\lambda}{\xi})))\,\xi^{-1}\,\bar\zeta
\;\big),
\end{array}
\end{equation*}
where $\lambda \in\bbR $ is parameter of the deformation.
\end{theorem}
\section{A variation of the central charge}
The outer superderivation $\ad\log\xi\in
H_0^1(\cS\Psi\cD\cO,\cS\Psi\cD\cO)$ defines a nontrivial 2-cocycle
with scalar values by the formula (\cite{AOR1})
\begin{equation}
\label{radul}
\widetilde C_{1}(A,B) = Str([\log\xi,A]\circ B).
\end{equation}

It is known that $\dim H^2(\cK(1),\bbC)=1$
(\cite{GLS}) and $H^2(\cK(1),\bbC)$ is spanned by the Neveu-Schwarz cocycle:
\begin{equation}
\label{leiters}
C(v_F,v_G)=\frac{-1}{4}\int_{S^{1|1}}F\eta^5(G)\;\vol(x, \theta)=
\frac{-1}{4}\int_{S^1}(4bd''+ac''')dx,
\end{equation}
where $F=a+2b\theta$ and $G=c+2d\theta$ with $a, b, c, d\in
C^\infty(S^1)$.
\begin{remark}
{\rm The restriction of the 2-cocycle (\ref{radul}) to the Lie
superalgebra $\cK(1)$ coincides with the Neveu-Schwarz cocycle.}
\end{remark}
\begin{proposition}
The restriction of the cocycle $\widetilde C_{1}$ to $\cK(1)
\hookrightarrow \cS\Psi\cD\cO_h$ with respect to the embedding
(\ref{map1}, \ref{explicite def} or \ref{map2}) is
\begin{equation}
\widetilde{\rho^\lambda}^*(\widetilde C_{1})=(h-4\lambda)C.
\end{equation}
\end{proposition}
\begin{proof}
By direct computations.
\end{proof}

\noindent{\bf Acknowledgements}
It is a pleasure to thank
Valentin Ovsienko who introduced us to the question of deforming
the Lie superalgebra of contact vector fields on $S^{1|1}$ inside
the Lie superalgebra of superpseudodifferential operators on
$S^{1|1}$. We also thank Claude Roger, M. Ben Ammar, B. Agrebaoui,
and  F. Ammar for helpful discussions.

\def\BibTeX{{\rm B\kern-.05em{\sc i\kern-.025em b}\kern-.08em
    T\kern-.1667em\lower.7ex\hbox{E}\kern-.125emX}}

\label{lastpage}
\end{document}